\documentclass[12pt]{article}
\usepackage{a4}
\usepackage{epsfig}
\newcommand{\beao}{\begin{eqnarray*}}
\newcommand{\eeao}{\end{eqnarray*}}
\newcommand{\be}{\begin{equation}}
\newcommand{\ee}{\end{equation}}
\newcommand{\bea}{\begin{eqnarray}}
\newcommand{\eea}{\end{eqnarray}}
\newcommand{\beq}{\begin{eqnarray}}
\newcommand{\eeq}{\end{eqnarray}}
\newcommand{\nn}{\nonumber}

\newcommand{\la}{\lambda}

\newcommand{\Ref}[1]{(\ref{#1})}

\begin{document}
\title{On the type of the temperature phase transition  in $O(N)$ models within a perturbative analysis  }
\author{M.~Bordag\thanks{e-mail: Michael.Bordag@itp.uni-leipzig.de}\\
{\small Universit\"{a}t Leipzig, Institute for Theoretical Physics} \\
{\small  Postfach 100 920, 04009  Leipzig, Germany} \\ [12pt]
{\small and} \\ [12pt]
V. Skalozub\thanks{e-mail: Skalozubv@daad-alumni.de}\\
{\small Dnipropetrovsk National University, 49010 Dnipropetrovsk, Ukraine}}
\date{}
\maketitle
\thispagestyle{empty}
\begin{abstract}
 We investigate  the type of the temperature phase transition  in the $N$ component $\la \phi^4$ ($O(N)$)
 model of scalar fields. Actual calculations are carried out in the beyond-super-daisy approximation (BSDA). The cases $N = 1$ and larger $N$ are considered separately.
  Using the solutions of gap equations we show that  the character of the phase
 transition depends on the account for graphs BSDA. The role of
 different kinds of diagrams (especially the "sunset" one)  is clarified. It is shown in a perturbation
 theory in the effective expansion parameter $N^{- 1/3}$ that the kind of the phase transition depends
 on the value of coupling $\la$. It turns from a weak  first-order to the second-order one
 for increasing $\la$. This is in agreement with the observation found recently for the $O(1)$ model in Monte Carlo
 simulations on a lattice.
  Comparison with  results of other authors is given.

\end{abstract}
\section{Introduction}
 Investigations of the temperature phase transition (PT) in the scalar  $N$ component $\la \phi^4 $ theories, called  $O(N)$ models,
have a long history. Various   approaches (perturbative, nonperturbative, analytic
numeric as well as Monte Carlo (MC) simulations on  a lattice) were used  and
 different results have been obtained (see, for instance, Ref.\cite{Zinn} and
 recent papers \cite{Marko}, \cite{Marko1} for references). A final conclusion was not settled, yet.  Nowadays, most calculations
advocate  a second order PT. However, a first order one is not excluded. In particular,
 using a partially resummed perturbative approach (resummation of two-particle irreducible
  (2PI) graphs) applied in the  beyond-super-daisy approximation (BSDA),
 in either $O(1)$ or $O(N)$ models, the weak first-order PT was seen in \cite{BS2001}-\cite{BS2002a}. Recently, in the $O(1)$ model it was shown in lattice MC calculations that the PT type  depends
on the value of coupling constant $\la$ . For $\la \leq 10^{-3}$ a first order
 PT was detected and for larger
$\la $ it converts into a second order one \cite{Bordag2010}.  This result
is unexpected because it is usually assumed  that the PT type is related with
 the analytic properties  of some order parameter in the neighborhood
  of critical temperature $T_c$. It is also known that near the PT
  temperature  the expansion parameter, which was initially small, is losing its smallness and perturbation theory
  in coupling constant becomes not reliable.   Hence one can conclude that noted result      requires further studies.

In recent papers \cite{Marko}, \cite{Marko1} the type of the
temperature PT in both  $O(1)$ and $O(N)$  models correspondingly
was investigated with accounting for a "basketball" (or "sunset")
diagram. It was stated that the PT becomes of second order just
due to this diagram contribution. In Ref.\cite{Marko1} a detailed
analysis was given by solving gap equations for particle masses
within two approximations  based on the 2PI resummations of
perturbation series. One of them called "hybrid approximation" is
an analytic one. These results and statements disagree with the ones
in our  earlier papers \cite{BS2001}-\cite{BS2002a}, where a
weak first-order PT was detected. So, it is of interest to find
out the origin of these discrepancies. This, in particular, is
important because of essential differences in approaches  applied
in these investigations. It is also desirable to find possible
causes for the dependence of the PT type on the value of $\la$ since
in Ref.\cite{Bordag2010} this behavior is the unexpected "fact of
calculations".

 In the present paper,  on the base  of
  Refs.\cite{BS2001}-\cite{BS2002a},  we analyze the corresponding gap equations,
  and solve them in different approximations for the BSDA graphs. One of
them is the sunset diagram investigated in
Refs.\cite{Marko},\cite{Marko1}.  In fact, this is only one term
in a series of the BSDA bubble chain diagrams taken into
consideration in Refs. \cite{BS2001} and \cite{BS2002a}. In these
papers  neither the role of the sunset diagram nor  reasons for
changing of the PT type were   investigated. Here, we consider
in detail its influence both, the
  unique  BSDA term  and a
particular element entering the complete series of diagrams.   For
simplicity of presentation, most calculations are carried out for
the $O(1)$ model where the change in the type of the PT is detected in
dependence on the kind of the BSDA diagrams taken into
account. In particular, we confirm the result on the second
order PT when the sunset diagram alone is accounted for. But this
behavior switches to first-order type when a more complete series of
  diagrams, the bubble chains, is taken into consideration.  Hence, the PT type is
controlled by the analytic structure of the resummed diagrams
independently of the value of $\la$. The case of $N \not = 1$ is
discussed in short. We show that  here the sunset diagram does not
change the type of the PT which is of weak-first-order at fixed
$N$.   We explain  the causes  of the discrepancies in the results
obtained in Refs.\cite{BS2001},\cite{BS2002} and \cite{Marko},
\cite{Marko1}. A change in the type of PT in dependence on the coupling
value was not observed.  To investigate this possibility, we apply
the perturbation theory in the effective expansion parameter $N^{-1/3}$ derived already in super daisy approximation (SDA) for large $N$
  \cite{BS2002}-\cite{BS2002a}. Within this analysis
  we find a possible cause for this phenomenon.

The paper is organized as follows. In the next section necessary
information  on the calculations carried out in
Refs.\cite{BS2001}-\cite{BS2002a} is adduced. Then we consider in
short the case of the $O(N)$ model and analyze the    solutions of the
gap equations in two cases, first, when the sunset diagram alone
is accounted for and second, when other terms of the complete bubble
chain series are taken into consideration. We show that in both
cases the weak first-order PT happens. We also explain the origin
of the discrepancies in the results obtained in Ref. \cite{BS2002}
and Ref. \cite{Marko1}.   In section 3 we consider the $O(1)$ model
and show  that in the case of one sunset diagram accounted for the
second order PT takes place.  However, it turns to the first order
one when the complete series of bubble chain diagrams is taken
into consideration. Section 4 is devoted to the investigation of the
PT type as function of the value of $\la$ in the limit $N \to \infty$ on
the base of the perturbation theory in the effective expansion
parameter $N^{- 1/3}. $ The dependence of the PT type is
determined.  The  last section is devoted to conclusions,
discussions  and comparisons with other approaches.

\section{PT in the BSDA for O(N) models}
The thermodynamical properties of the model are
described by the partition function
\begin{equation}\label{gen-func}
Z=\int D\varphi~e^{-S[\varphi]},
\end{equation}
where $\varphi$ is $N$ component real scalar field, and the action is
\begin{equation}
S =\int
dx\left(\frac12\partial_\mu\varphi(x)\partial_\mu\varphi(x)-\frac12m^2\varphi(x)^2+\frac{\lambda}{4 N}(\varphi(x)^2)^2 \right).
\end{equation}
Here, $\phi^2(x) = \sum_{i=1}^N \phi^i(x) \phi^i(x)$. More details on the model see in
Refs.\cite{BS2001}-\cite{BS2002a}.

We use the second Legendre transform representation
\be \label{Ltransform} W = S[0] + \frac{1}{2} Tr log \beta - \frac{1}{2} Tr \Delta^{-1} \beta + W_2[\beta], \ee
where $W_2[\beta]$ is the sum of all 1PI diagrams with propagators $\beta$.
Further, $\Delta^{-1} = - \partial^2_\nu + m^2  $ is the  inverse free propagator in Euclidian space and $\beta$ is exact the propagator subject to the
Schwinger-Dyson equation
\be \label{SDe} \beta^{- 1}(p)  = \Delta^{-1} - \Sigma[\beta] (p) \ee
with $\Sigma[\beta] (p) = 2 \frac{\delta W_2}{\delta \beta (p)}$, which is the sum of all two-particle irreducible (2PI) self-energy graphs with   propagators $\beta$. In the imaginary time formalism the operator $"Tr"$ is
\be \label{Tr} Tr_p = T \sum\limits_{l = - \infty}^{\infty} \int \frac{d^3 p}{(2 \pi)^3} \ee
with the 4-momentum $p = (2 \pi l T, \vec{p})$. We indicate the functional argument $\beta$ by square brackets and momentum by round ones.

Now let us remind our approximations. In Ref.\cite{BS2002},
Eqs.(45), (55),  the most general gap equations for Higgs, $\eta$,
and Goldstone, $\phi$, fields have been derived in the extremum
of the free energy functional taken in the 2PI approximation which
accounts for in the BSDA part an infinite series of bubble chain
diagrams. We write down them here in the form
\be \label{Metagap1} M^2_\eta = 2 \tilde{\Gamma}  \left( m^2 - \frac{3
\la}{N} \Sigma^{(0)}_\eta - \la \frac{N - 1}{N} \Sigma^{(0)}_\phi + 2
\Gamma \right)\ee
for  the Higgs field mass $M_\eta $ and
\be \label{Mphigap1} M^2_\phi = \frac{2 \la}{N} \left(\Sigma^{(0)}_\phi
- \Sigma^{(0)}_\eta\right) - g M^2_\eta + f(\eta, \phi) \ee
for the Goldstone field mass $M_\phi $. The functions $\Gamma,
\tilde{\Gamma}, \Sigma^{(0)}_\eta, \Sigma^{(0)}_\phi, f(\eta,
\phi)$ are calculated in Eqs.(46), (50), (53) Ref.\cite{BS2002}.
Eq.\Ref{Metagap1} at $N = 1$ coincides with the gap equation for
the $O(1)$ model in Ref.\cite{BS2001}. These equations are derived
from renormalized special series graphs entering 2PI free energy.
The renormalization is fulfilled by the counter terms obtained at
zero temperature. For more details see
Refs.\cite{Kopper2000},\cite{Bordag2002ab}.

Solving of the gap equations
in the extremum of free energy is the key point of the  approach
used. Due to this choice, the physical masses of the finite temperature excitations  are determined
as the locations of the poles.
Next what is important, as the consequence of this condition the
term,
\be \label{asymterm} f(\eta, \phi) = \frac{\delta D}{\delta
\eta(0)} - \frac{1}{N - 1} \frac{\delta D}{\delta \phi(0)}, \ee
with indefinite sign has appeared in Eq.\Ref{Mphigap1}.  In the $O(N)$ models, this is
the main reason for  first-order PT (see below).

Now, let us write down the expressions in Eqs.\Ref{Metagap1}-\Ref{asymterm}
 in the approximation when only the first terms
(just the sunset contributions as in Ref.\cite{Marko}) from  the
series of all diagrams derived in the Appendix of \cite{BS2002}
are taken into account. We have (see Ref.\cite{BS2002})
\be \label{Gamma11} \Gamma^{(1)} =
- \frac{6 \la}{N} Tr_q [\beta_\eta (\frac{a}{6} + \frac{\epsilon
b}{2})] = \frac{\la^2}{N} (\frac{3}{N}  S_\eta + \frac{N - 1}{N}
S_{\eta \phi_2} ), \ee
and
\be \label{Mphigap} M^2_\phi = \frac{2 \la}{N} (\Sigma^{(0)}_\phi
- \Sigma^{(0)}_\eta) - g M^2_\eta + f(\eta, \phi) \ee
for the Goldstone field mass
\be \label{dDf} \frac{\delta D^{(1)}}{\delta \phi(0)} =
\frac{\la^2}{3 N} [ (1 + 1/N) (N + 1 + 2 \frac{N - 2}{N + 1})
S_\phi + 3 \frac{N - 1}{N} S_{\phi \eta_2}], \ee
where $\Gamma^{(1)}, D^{(1)}$ denote that only one term of $a$ or
$b$ types is taken into consideration and we introduced the
notations for the sunset diagrams
\bea \label{Si} S_{\eta,\phi} &=& \frac{T^2}{32 \pi^2} \left(1 - 2 Log
\left[\frac{3 M_{\eta, \phi}}{\mu}\right]\right), \\ \nn
  S_{\eta,\phi_2} &=& \frac{T^2}{32 \pi^2} \left(1 - 2 Log \left[\frac{ M_{ \eta} + 2 M_{\phi}}{\mu}\right]\right), \\ \nn
S_{\phi,\eta_2} &=& \frac{T^2}{32 \pi^2} \left(1 - 2 Log \left[\frac{2
M_{\eta} + M_{\phi}}{\mu}\right]\right). \eea
Here, $\mu$ is a normalization scale. Calculation of these contributions by different methods  see, for example,in Refs. \cite{Arnold1994}-\cite{Shaposhnikov1994}. In Eq.\Ref{Mphigap} the notation $\Sigma^{(0)}= Tr_q \beta (q)$ is used, where $\beta $ is determined below in Eq.\Ref{propagator}.

Obviously,   for small $M_\phi$, in Eq.\Ref{asymterm} in the
difference  the signs of the leading log-terms $Log [\frac{3
M_{\phi}}{\mu}] $ are dominant. This is because, as we showed in
the SDA \cite{BS2002}, the mass $M_\eta $ is always larger than $M_\phi$. Just
these $M_\phi$-terms control the type of the phase transition at
high temperature when $M_\phi$ is small.

As we see from Eq.\Ref{Gamma11}, there are no $Log [\frac{3
M_{\phi}}{\mu}] $ terms in $\Gamma^{(1)}$. In this limit, in
Eq.\Ref{dDf} the first term $S_\phi$ is dominant. Also, we see
that $Log [\frac{3 M_{\phi}}{\mu}]$ enters the function $f(\eta,
\phi)$ in Eq.\Ref{asymterm} with the sign "plus". So, it gives a
negative contribution in the gap equation \Ref{Mphigap}.
This behavior is sufficient to generate the first order PT.
 Really, due to the negative sign noted there are two
crossing of the parabola in the l.h.s. with the behavior in
the r.h.s. One crossing is at small $M_\phi$ and other one at
larger $M_\phi$. As a result, substituting these values into
 Eq.\Ref{Metagap1}
we have to obtain two solutions for $M_\eta$. This is because in
the $\Gamma$ the term $Log [\frac{3 M_{\eta}}{\mu}] $ enters with
the negative sign. So that at small $M_\eta$ this contribution is
positive and only one crossing of the l.h.s. with the r.h.s. is
possible. In this way the first order phase transition is
realized. Note that this is just due to the properties of the
sunset diagrams, in contrast to what is stated  in
Ref.\cite{Marko}. These authors erroneously related the first order type of the PT
observed in Ref.\cite{BS2002} with the ansatz for the propagators
\be \label{propagator} \beta^{-1}_{\eta,\phi}(p) = p^2 + M^2_{\eta,\phi} \ee
used therein. However, this is not the case and actual cause is the solution of the gap equations
in the extremum of free energy. Note also that the behavior of the $\tilde{\Gamma}$ in Eq.\Ref{Metagap1}
is not important, it does not lead to new crossings.

 We have to note once again that the
term with indefinite sign in Eq.\Ref{asymterm} is the consequence of  passing to the
extremum of free energy.  In contrast, in Ref.\cite{Marko1} the
gap equations (Eqs.(6), (7) or (38), (39)) were solved out of the extremum of the
free energy. So,  the sunset contributions $S_\eta$ and
$S_\phi$ into the effective potential  (Eq.(14), \cite{Marko1})
enter  with the same signs.  This is the reason why the second
order PT was found. More details on this point are given in the
last section.

As general conclusion of the above analysis we note that in the
analytic calculations accounting for the sunset diagram alone the
weak-first-order PT follows if the gap equations \Ref{Metagap1},
\Ref{Mphigap1} are solved in the extremum of free energy, that is
if the condition $\partial W /\partial v^2 = 0$, where $W$ is free
energy and $v$ is vacuum value of the Higgs field,  is used
explicitly.
\section{Phase transition in $O(1)$ model}
To investigate  the  $O(1)$ model we set N = 1 in
Eq.\Ref{Metagap1} that omits the contribution of the $\phi$
fields.  The structure of the functions $\Gamma$ and
$\tilde{\Gamma}$ was presented in Ref.\cite{BS2001}.

The notations are:
\be \label{Gamma} \Gamma = Tr_q \beta(q) \Sigma_1(q) \frac{1 - 6 \la \Sigma_1(q)}{1 + 3 \la \Sigma_1(q)},\ee
(Eq.(18), \cite{BS2001})
and
\be \label{Gamma1} \tilde{\Gamma} = 1 - 3 \frac{\delta \Gamma}{\delta \beta(0)},\ee
(Eq.(25), \cite{BS2001}), where $\beta$ is given by
Eq.\Ref{propagator} and
\be \label{Sigma1} \Sigma_1(p) = Tr_q \beta(q)\beta(q + p),\ee
(Eq.(20), \cite{BS2001}).

In order to be in complete correspondence with Ref.\cite{BS2001} we rewrite Eq.\Ref{Metagap1} in the form
\be \label{ge1} M^2 = 2 \tilde{\Gamma}  \left( m^2 - \frac{3
\la}{N} \Sigma^{(0)} + 6 \la^2 \Gamma \right),\ee
where $\Sigma^{(0)} = Tr_q \beta(q)$ and $m$ is the zero temperature mass. This equation is also written in the extremum of free energy.

Now, we consider the approximation when among the BSDA terms the
sunset diagram only is taken into consideration. In Eqs.
\Ref{Gamma},  \Ref{Gamma1} this means that we have to put $\la =
0$. As a result we have
\be \label{Gammass} \Gamma^{ss} = Tr_q [ \beta(q) \Sigma_1(q)] = S_\eta,~~ \tilde{\Gamma}^{ss} = 1,\ee
where $S_\eta$  is given in Eq.\Ref{Si}.

We solve Eq.\Ref{ge1} graphically by presenting the l.h.s. and the
r.h.s. in one plot. As the expression for $\Sigma^{(0)}(M)$ the
high temperature expansion adduced in Eq.(55) of \cite{BS2001} is
used:
\be \label{Sigma0T} \Sigma^{(0)}(M) = \frac{T^2}{12} - \frac{M T}{4 \pi} + \frac{1}{16 \pi^2} \left(M^2\left(\ln \frac{8 \pi^2 T^2}{m^2} - 2 \gamma\right) + 2 m^2\right) + O(M^2). \ee
Here, $\gamma$ is Euler's gamma. To show graphical solutions we
introduce the dimensionless variables $\tau = \frac{T}{m}$ and
$\rho = \frac{M}{m}$  and  set the normalization scale $\mu = m$
in $S_\eta$, although  this is not essential.

In Figs.1-2 we show the results for two temperatures. Along the OX axis, the values of the scalar field mass $\rho$ are measured.
 Along the OY axis,  the values of the l.h.s. and the r.h.s. are present. As we see,
  there is only one crossing of the curves at some point $\rho = \rho(\tau)$. With increasing of the temperature
  the crossing point is shifted to smaller masses.  This means the second order PT and is
   in agreement with the result in Ref.\cite{Marko}.
\begin{figure}
\begin{center}
\includegraphics[bb=0 0 260 162,width=0.8\textwidth]{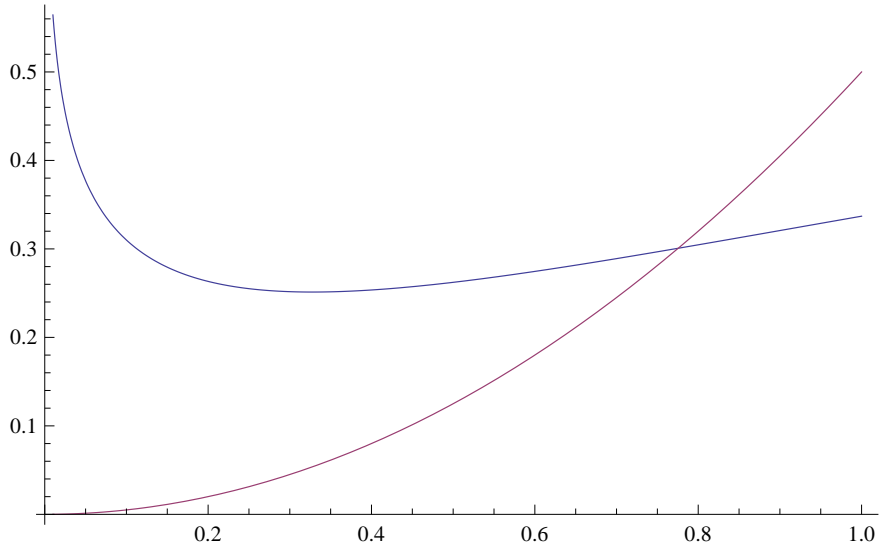}
\end{center}
\caption{Solution of gap equation for $\la = 0.915$ and $\tau = 2$ with the sunset diagram taken into consideration, notations are given in the text. }
\end{figure}
~\\~
\begin{figure}
\begin{center}
\includegraphics[bb=0 0 260 154,width=0.8\textwidth]{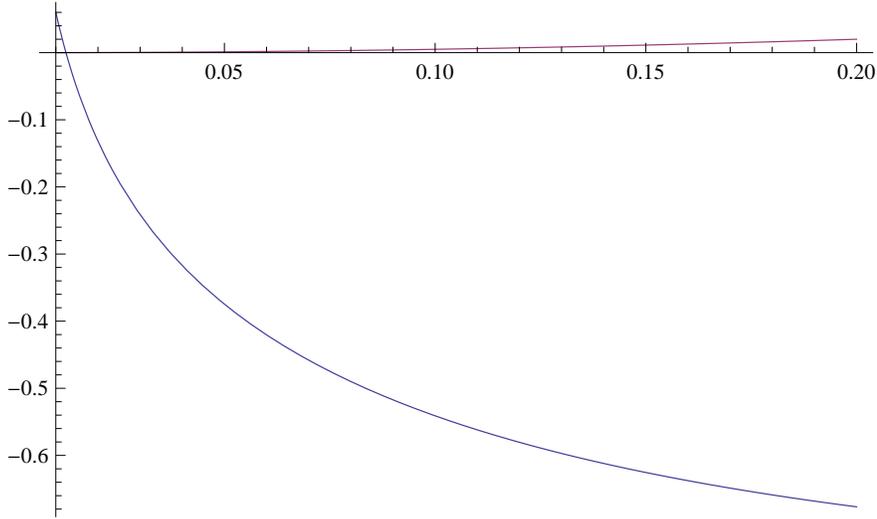}
\end{center}
\caption{Solution of gap equation for $\la = 0.915$ and $\tau = 3$ with the sunset diagram taken into consideration, notations are given in the text. }
\end{figure}

However, the sunset diagram is only one term in the  bubble chain
series resulting in  $\Gamma$, Eq.\Ref{Gamma}. Since the
numerator in this expression is not a function with definite sign, the
value of $\Gamma$ may change in dependence on the behavior of $\Sigma_1$.
In Ref.\cite{BS2001}, Eq.(57), the high temperature expansion for
this function is adduced:
\be \label{eq57} \Sigma_1(0) = \frac{T}{4 \pi M} + O(1). \ee
Next, let us investigate the  gap equation \Ref{ge1} with
accounting for the full expression for $\Gamma$, Eq.\Ref{Gamma},
with this $\Sigma_1$, as well as the complete $\tilde{\Gamma}$,
Eq.\Ref{Gamma1}. Corresponding solutions are shown in Figs.3-4.
For this case, there are two crossings of the curves - one at small
values of $\rho = \rho(\tau)$ and another at larger ones. This
 is typical behavior signalling a first order PT. Crossing at small $\rho$ corresponds to the maximum of free energy
 at a given temperature. Crossing at larger $\rho(\tau)$ determines the value of the scalar particle mass in the minimum
 of the effective potential, that is, the physical mass of the excitation. Note here that the behavior of $\tilde{\Gamma}$
 is not very essential. It influences a little bit the parameters of the PT. The type of the PT is completely determined
 by the  function $\Gamma$. As a result,  the first order PT is observed. Hence it follows that
  not only the sunset diagram but also all the other terms of the series are important and control  the type of the PT.
   As concerns the dependence of the PT type on the value of $\la$, we do not see that in the given approximation for a wide
    interval of $\la$. Typical behavior is depicted in the above plots. Changing  of the PT type follows in dependence
    on the expression for $\Gamma$, only.  Note also that  the choice of the truncated  functions  can influence the type
     of the PT. For example,  in Ref.\cite{BS2001} more rough  approximations for $\Gamma$  were used. As a result,
     weak-first-order  PT followed and a second order one was not    observed.
\begin{figure}
\begin{center}
\includegraphics[bb=0 0 260 154,width=0.8\textwidth]{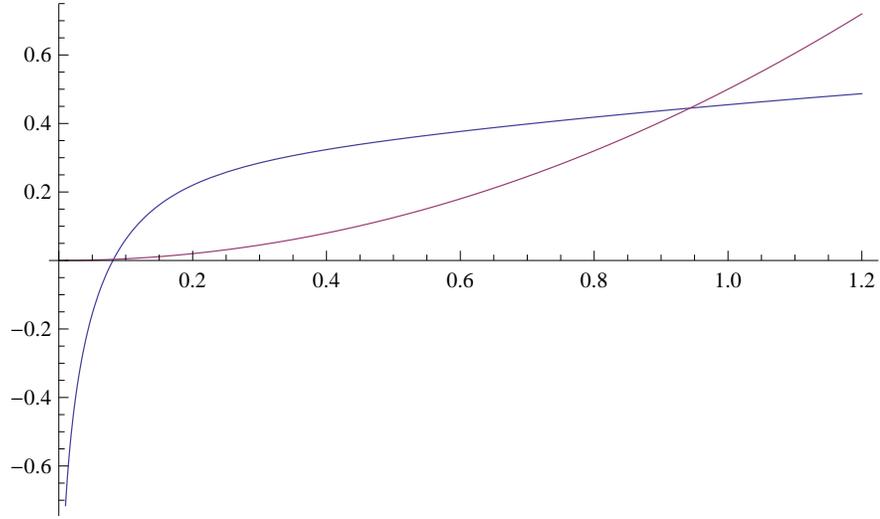}
\end{center}
\caption{Solution of gap equation for $\la = 1$ and $\tau = 1.8$ with the complete set of bubble chain diagrams included, notations are given in the text. }
\end{figure}
~\\~
\begin{figure}
\begin{center}
\includegraphics[bb=0 0 260 153,width=0.8\textwidth]{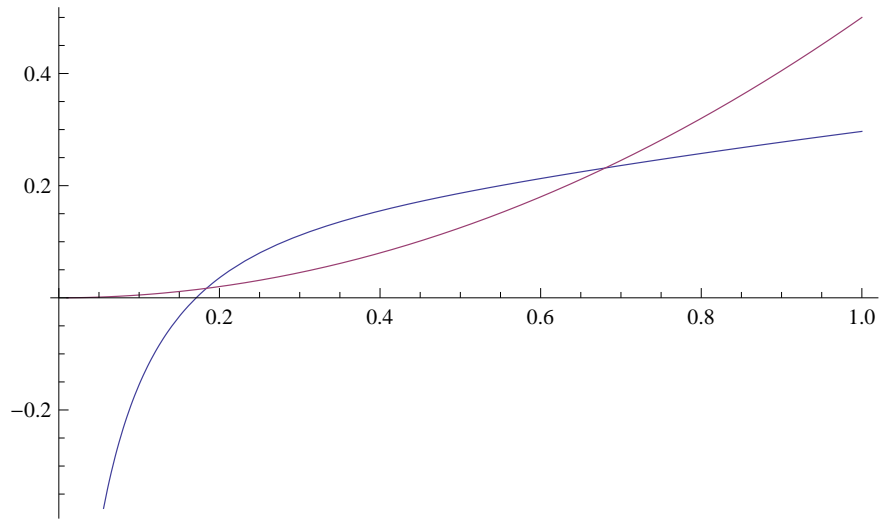}
\end{center}
\caption{Solution of gap equation for $\la = 1$ and $\tau = 2$ with the complete set of bubble chain diagrams included, notations are given in the text. }
\end{figure}

\section{$\la$-dependence  in the   $\frac{1}{N^{1/3}}$ expansion}
In this section we  investigate the  $\la$ dependence of the PT type within the perturbation theory based on the  effective expansion parameter $N^{-1/3}$ near the PT temperature \cite{BS2002}, \cite{BS2002a}.

First, let us adduce information necessary for what follows. This  perturbation theory accounts  for the contributions of the BSDA graphs or series of graphs. It is based on the particle masses $M_\eta^{(0)}$ and $M_\phi^{(0)}$ obtained as solutions to the gap equations which  take into consideration  the SDA graphs, only.  That is, we have to omit  the last terms in Eqs.\Ref{Metagap1}, \Ref{Mphigap1} and obtain (see Eq.(17) in \cite{Bordag2002ab}):
\bea M_\eta^{(0)} &=& \frac{\la T_+}{4 \pi} \left(\frac{1}{(2 N)^{1/3}} -\frac{1}{2 N} + \cdot \cdot \cdot \right), \nn\\
M_\phi^{(0)} &=& \frac{\la T_+}{2 \pi} \left(\frac{1}{(2 N)^{2/3}} -\frac{1}{2 N} + \cdot \cdot \cdot  \right). \label{masses0}\eea
These expressions were calculated in the limit $N \to \infty$, $T_+$ is the  upper spinodal  temperature. The choice of  $T_+$ is motivated by simplicity  of analytic expressions for this case.

The main steps of calculations are as follow. Each BSDA diagram containing $n$ vertexes comes with the factor $(\frac{\la}{N})^n$ and has to be written with the masses \Ref{masses0}. Then in the imaginary time formalism one must shift the three momentum of a loop $\vec{p} \to M \vec{p}$ and the temperature $T \to \frac{T}{M} \sim T N^{2/3}$, where for definiteness we have substituted the mass  $ M_\phi^{(0)}$. After these shifts the common $N$-dependent factor of the diagram is obtained. Hence,  we can see that the high temperature approximation is well applicable for large $N$. As a result, the contribution of the static mode $l = 0$ in the Matsubara sum is dominant. As it is known, in the static limit ($l = 0$) a theory becomes effectively three dimensional.
So, this approximation is sufficient for solving most problems of interest.
This, in particular, means that we can calculate corrections to the SDA results in the three space dimensional $O(N)$ model.

 In Ref.\cite{Bordag2002ab}, as application, an  infinite series of the bubble chain diagrams with the $ M_\phi^{(0)}$ mass was calculated at the $T_+$ temperature. Remind, in the SDA approximation a weak-first-order PT was determined \cite{BS2002}. In this section, to investigate the $\la$-dependence of the PT, we recalculate these corrections in the $d = 3 ~~O(N)$ model  and check how the value of $\la$ influences the difference between the lower $T_-$ and  the upper $T_+$ spinodal temperatures.

To realize  that we use the result of summing up the series of the bubble chain diagrams in the restored phase (Eq.(26) in \cite{Bordag2002ab}):
\bea D_\phi &=& - \frac{\la^1}{N^2} Tr_p \log\left(1 + \Sigma_\phi^{(1)}(p) N^{2/3}\right)\nn \\
&+&  \frac{\la^1}{N^{4/3}} Tr_p  \Sigma_\phi^{(1)}(p) - \frac{\la^1}{4 N^{2/3}} Tr_p  \left(\Sigma_\phi^{(1)}(p)\right)^2.\label{Dphi}\eea
Here, $\Sigma_\phi^{(1)}(p)$ is defined in Eq.\Ref{Sigma1} and $Tr_p = \Sigma_{l=-\infty}^{l=+\infty} \int\limits_{-\infty}^{+\infty} d^3 p $ .
For simplicity, we restrict ourselves to the large $N$ case and take into consideration the last term, only. Other terms give next-to-leading corrections.

Now, we have to renormalize this contribution. In
Ref.\cite{Bordag2002ab} a general renormalization procedure   was
developed. In particular, it was shown for the vertexes and bubble
chain graphs that the counter terms removing divergencies at zero
temperature are sufficient for doing that  at finite temperature,
as it should be.

Here, we consider the renormalization of the last term in Eq.\Ref{Dphi} in the three space dimensions.
We present the term of interest in the equivalent form
\be \label{Dphi1} D_\phi^{(3)} = - \frac{\la^1}{ N^{2/3}}\frac{T }{(2 \pi)^3 } \int d^3 p\beta(p,M_\phi) S_\phi(p, M_\phi), \ee
where
\be \label{Sphi} S_\phi(\vec{p}^{\ 2}, M_\phi) = \frac{T^2 }{(2 \pi)^6 } \int d^3 k d^3 q \ \beta(k,M_\phi)\beta(q - k,M_\phi)\beta(q - p,M_\phi)  \,  \ee
is the sunset diagram, and we set $l = 0$ in the Matsubara sums. This representation is convenient because we need it in the functional derivative  $\delta D_\phi/\delta \beta$ as in Eq.\Ref{asymterm}. The function  \Ref{Sphi} was analytically calculated  in Minkowski space-time   for arbitrary $d$ in terms of Hypergeometric functions  \cite{Tarasov2006}. It is not difficult to transform that expression for our case.   More early calculations of the sunset diagram see in Refs.\cite{Arnold1994}-\cite{Shaposhnikov1994}. So, the counter term for this diagram $C_3$ can be easily found. As we see from the above expression, the integral is logarithmical divergent and so the counter term is a simple pole in the expansion over the deviation $2 \epsilon = 3 - d$ in the limit $\epsilon \to 0$. It has to be subtracted from the $S_\phi(\vec{p}^2, M_\phi)$  to get a finite part.  To obtain the renormalized $D_\phi^{(3)}$ we have to substitute the renormalized expression \Ref{Sphi} in Eq.\Ref{Dphi1} and introduce a counter term which cancels the divergent part coming from  the  p-integration.  This is  standard procedure. In fact, the last step is not necessary  for what follows because here we need in the contribution  of the sunset diagram taken at $p = 0$.

Then we proceed as in Ref.\cite{Bordag2002ab} and solve the gap equations \Ref{Metagap1}, \Ref{Mphigap1} perturbatively. With the only term $D_\phi^{(3)}$  taken into account we have
\be \label{Metagap2} M^2_\eta =  m^2 - \frac{3
\la}{N} \Sigma^{(0)}_\eta - \la \frac{N - 1}{N} \Sigma^{(0)}_\phi \ee
 and
\be \label{Mphigap2} M^2_\phi = \frac{2 \la}{N} \left(\Sigma^{(0)}_\phi
- \Sigma^{(0)}_\eta\right) + f(M_\phi), \ee
where now in the large $N$ limit
\be \label{f2} f(M_\phi) = - \frac{2}{N - 1} \frac{\delta D_\phi^{(3)}}{\delta \beta_\phi(0)}
= \frac{\la}{N^{5/3}} \frac{T^2}{32 \pi^2} \left(1 - 2 \log \frac{3 M_\phi}{\mu} \right).\ee
 For the contributions $\Sigma^{(0)}_\eta$ and $\Sigma^{(0)}_\phi$ we use the leading terms in the high temperature expansion \Ref{Sigma0T}
\be \label{Sigma0T2} \Sigma^{(0)}(M_{\eta,\phi}) = \frac{T^2}{12} - \frac{M_{\eta,\phi} T}{4 \pi}+ \cdot \cdot \cdot  .\ee
To obtain perturbative solutions we write
\bea \label{masses2} M_{\eta} &=& M_{\eta}^{(0)} + x, \\ \nn
M_{\phi} &=& M_{\phi}^{(0)} + y , \eea
where the masses \Ref{masses0}  are substituted and $x, y$ are small corrections which must be calculated.
The result  is as follows \cite{Bordag2002ab}:
\bea \label{xy} x &=& \frac{1}{3} \frac{T_+}{32 \pi} \frac{1}{N^{2/3}} \Bigl(1 - 2 \log \frac{3 M_\phi^{(0)}}{\mu} \Bigr), \\ \nn
y &=& \frac{1}{2}  \frac{2^{2/3}}{3^{1/2}} \frac{T_+}{32 \pi} \frac{1}{N} \Bigl(1 - 2 \log \frac{3 M_\phi^{(0)}}{\mu} \Bigr). \eea
The values and  signs of corrections depend on the choice  of
normalization parameter $\mu$. This situation is known in field
theory and  particle physics. In the context of temperature PT in
the standard model the choice of $\mu$ is discussed in
Refs.\cite{Shaposhnikov1994},\cite{Shaposhnikov1994lat}. In
Appendix B of the former paper some details of renormalization in
$d = 3$ space dimensions relevant to our case are considered. This
choice  is also discussed in Ref.\cite{Marko1}. Although the
result has to be independent of $\mu$, in practice,  in
perturbation calculations $\mu$ is chosen to be of the order of
typical mass in the problem. This removes large logarithmic
corrections. In our case, we set $\mu = M_\phi^{(0)}$ because this
is particle mass \Ref{masses0} at close to the $T_c$  temperatures.  With
this choice, we see that the function \Ref{f2} and corrections
\Ref{xy} have negative signs that  decreases the masses of
particles. Note also that in section 3 we set $\mu = m$. This
choice was inessential because we were interested in qualitative
behavior  which is determined by the number of crossing of the
curves presenting l.h.s. and r.h.s. of equations.

Similarly, the mass correction can be calculated in the restored
phase where  $M_\phi = M_\eta = M_r$.  In Ref.\cite{Bordag2002ab}
it is shown that if we present the mass as $M_r = M_r^{(0)} + z$,
the correction $z$ is given by the expression (see Eq.(52)in
 \cite{Bordag2002ab}):
\be \label{z} z = \frac{ f(M_r^{(0)})}{ 2 M_r^{(0)} + \frac{\la T}{4 \pi}}.\ee
This is  negatively  valued function. As a result, the mass in the restored phase is also decreased.
The mass $M_r^{(0)}$ was calculated already in the SDA in Refs.\cite{BS2002}, \cite{BS2002a}. It looks as follows:
\be \label{Mr0} M_r^{(0)} = - \frac{\la (N + 2) T}{8 \pi N} + \sqrt{\Bigl(\frac{\la (N + 2) T}{8 \pi N}\Bigr)^2 - m^2 + \frac{\la (2 + N) T^2}{12 N}}.\ee
Clearly that the mass   should be positive parameter.

In SDA, the temperature $T_-^{(0)}$ is determined from the condition $M_r^{(0)}= 0$. Its value is easily calculated with the expression \Ref{Mr0}:
 \be \label{T-} T_-^{(0)} = \frac{2 m}{\sqrt{\la}} \frac{\sqrt{3} \sqrt{ N}}{ \sqrt{2 + N}} .\ee
 Corresponding value for $M_r^{}= 0$ can be found from similar equation $M_r^{}= M_r^{(0)}+ z = 0$. To
 estimate the influence of the $z$ correction, we consider   numerical  examples. Let us  introduce dimensionless variables: $\tau = \frac{T}{m}$, $m_r^{(0)} = \frac{M_r^{(0)}}{m}$ and calculate
$\tau_-^{(0)}$  and $\tau_-^{z}$ for $\la = 1, 5$. For definiteness we also take $N = 8$. We obtain for $\la = 1$, $\tau_-^{(0)}= 12$ and $\tau_-^{z}= 12.208$. For $\la = 5$ these numbers are $\tau_-^{(0)}= 2.4$ and $\tau_-^{z}= 2.408$. Hence, it follows that $z$-correction slightly increases the lower spinodal temperature. The same, of course, takes place for other values of the coupling and number of components $N$.

Now consider the  temperature $T_+$. In the SDA it was found  that the relation holds \cite{BS2002}, \cite{BS2002a},\cite{Bordag2002ab}:
\be \label{T+T-} T_+^{(0)} = T_-^{(0)}\left(1 + \frac{9 \la}{16 \pi^2} \frac{1}{(2 N)^{2/3}} \right). \ee
As we see, for a given $N$ the difference between these
temperatures  is increasing function of $\la$. So, in this
approximation, the  larger is $\la$ the stronger is first order
PT.

The BSDA correction to the $T_+^{(0)}$ was also calculated. The
result is as follows (Eq.(61) in \cite{Bordag2002ab}),
\be \label{T+cor} T_+ = T_+^{(0)} \left(1 + \frac{9 \la}{32 \pi^2} \frac{(1 - 2 log \frac{3 M_\phi^{(0)}}{\mu})}{N^{5/3}} \right).\ee
Here, $T_+^{(0)}$ is given in Eqs.\Ref{T-},\Ref{T+T-} and for
$\mu$ we have to substitute the value $M_\phi^{(0)}$ as before.
Since the value $1 - 2 \log 3 = -1.1972$ is negative, we observe a
nontrivial  behavior of the r.h.s. as a function of $\la$.  It
follows from the product of two brackets standing in
Eqs.\Ref{T+T-} and \Ref{T+cor}. The former one is increasing and
the latter - decreasing function of  $\la$. For a fixed $N$ and
sufficiently large $\la$,  the brackets in Eq.\Ref{T+cor} becomes
enough  small that turns $T_+^{}$ to go down. As a result, the
difference between $T_+$ and $T_-$ is decreased. The
weak-first-order PT derived in the SDA presents  a tendency for
converting  into the second order one. Just such type  behavior
was observed for the $O(1)$ model on a lattice  \cite{Bordag2010}.
From Eqs.\Ref{T+T-} and \Ref{T+cor} it also follows that the small
parameter has to be  $\frac{\la}{16 \pi^2} \frac{1}{(2 N)^{2/3}} $
and therefore the value of $\la$ can be sufficiently large and
dependent on $N$. For example, for $N = 2$, the parameter is small
for $\la$ values   10-20.

Thus, close to the temperature $T_c$, the perturbation theory in
the effective parameter $N^{-1/3}$  gives a possibility for
determining  causes for changing the PT type dependently
on the coupling value. It consists in specific $\la$-dependence of the
sunset diagram contributions to $T_+$. Since in three dimensions
this is the only BSDA diagram with such type dependence, the
importance of this contribution is clarified.
\section{Conclusion}
We have analyzed the temperature PT in the $O(1)$ and $O(N)$
scalar field models.  Actual calculations were carried out in the
BSDA  for 2PI free energy functional. The gap equations were
solved graphically within a high temperature approximation for
Green's functions.

In the $O(1)$ model, we observed the change of the PT type
in dependence on the BSDA  diagrams taken into consideration. More
definitely, for the case of the sunset diagram alone we found a
second order PT, that is in agreement with the results of
Ref.\cite{Marko}. However, for the complete bubble chain series,
containing the sunset one as a particular element, a first order
PT was detected.  We showed that the change of the PT type is
related with the  structure of the $\Gamma$ term, Eq.\Ref{Gamma},
where the sunset diagram stands as a common factor. So, the
properties of this diagram were taken into account completely. We
can  conclude  that the sunset diagram  does not uniquely control
the type of the  PT, other contributions are also important.

In Refs.\cite{Marko}, \cite{Marko1} it is, in particular,  stated
that the first order PT detected  in Ref.\cite{BS2001} is the
consequence of the ansatz for the full  propagator $\beta$
Eq.\Ref{propagator}. However, as we showed  above, this is not the
case and the actual cause is  a rough approximation for the
$\Gamma$ function Eq.\Ref{Gamma} used in Ref.\cite{BS2001}. It is
worth   mentioning that, as it is demonstrated  in
Ref.\cite{BS2002}, the SDA possesses a lot of attractive
properties. One of them  is a possibility to derive an effective
expansion parameter $\sim N^{-1/3}$ near the PT temperature
 for large $N$.  This allows  to construct
perturbation theory in this parameter. We  have shown in the BSDA
that in the $O(N)$ models the temperature PT is a
weak-first-order. This also disagrees with the results obtained in
Ref.\cite{Marko1}, where  the sunset diagram only was taken into
account and a second order PT was  observed. In section 2, we have
demonstrated  that in this case  the first order PT follows from
solving of the gap equations in the extremum of free energy. In
fact, this is the main reason for the discrepancies in the
obtained results.  As we mentioned in the Introduction, solving of
gap equations in the extremum of free energy is the key element of
the approach applied in Refs.\cite{BS2001}-\cite{BS2002a}. This
requirement is based on the general principle of thermodynamics
stated  that  physical masses  have to be obtained in the minimum
of  free energy.  In Eqs.\Ref{Metagap1},\Ref{Mphigap1}
this requirement is satisfied by construction.

Let us continue our  comparison with the approach applied in
Refs.\cite{Marko}, \cite{Marko1}.  Two points of it
may result in the discrepancies. First is the renormalization
procedures used. In our consideration, a standard renormalization
with zero temperature counter terms is applied, as it is presented in
the previous section and
in Refs.\cite{BS2001},\cite{Bordag2002ab}.  Moreover, for large N the static mode contribution is  well
applicable and the theory is effectively three dimensional one.
 In contrast, in Refs.\cite{Marko}, \cite{Marko1}
renormalization is fulfilled at finite temperature with
temperature dependent counter terms. As a
consequence, uncontrolled finite temperature contributions
affect  the results. Second, in our method the solution to gap
equations in the minimum of free energy results in the pole
physical masses of particles. Just due to this choice the
term \Ref{asymterm} with indefinite appears and generates
first-order PT. In Refs.\cite{Marko}, \cite{Marko1} the particle
masses are calculated in two steps. At first, the gap equations are
solved at arbitrary vacuum condensate giving the pole masses which
are then substituted in the effective potential. Next by
minimization of the potential and calculating its second derivative in the
minimum the "curvature masses" are calculated. They are considered as
 the physical  masses. In the course of these calculations,
the sunset diagrams enter the stationary
equation with the same signs.  This is formal mathematical
cause of the second order PT. We also remind  that the solution of gap equations
corresponds to the summation of the infinite series of loop diagrams that is a nonperturbative
result. So,  one joins  these  results with the minimization of the two-loop effective potential.
This procedure does not coincide with solving of gap equations in the minimum of free energy.
So, the results could be different.
Besides, the usage of the two sorts of  masses makes this calculation
 procedure complicated.  We believe  that pole masses
are close related to the particle spectrum and should be found in the minimum
of the effective potential not vice versa.

As concerns the dependence of the PT type on the coupling value
observed in MC  simulations on the lattice \cite{Bordag2010}, it
was not detected in analytic calculations within the solutions to
gap equations fulfilled in the BSDA. In this approach, the PT type
is determined by the number of crossings of the curves depicted in
plots Figs. 1-4. This number is strictly related with the signs of
the log-terms entering the dominant sunset diagram contributions.
If this sign is positive at small $M$, the only one crossing
happens  and the second order PT follows. Otherwise the first
order PT takes place. These properties are independent  of the
$\la$ value.  However, different kind resummations can be carried out
that makes this sign  out of any control in general.

It is also interesting to compare our results with the ones derived in analytic
solutions of exact renormalization group equations \cite{Tetradis1996}.  In this approach,
in the framework of the effective average action, the analytic solutions for the scale dependence
of the potential have been obtained for  O(N) models in different space dimensions.
Actual calculations were carried out in the limit $N \to \infty$.
For  d = 4, a second order PT has been found. This is in agreement with our results for large
N. Here, it is interesting to note that for some relations between the parameters of the models
 a first order PT was observed. This behavior is similar to the PT $\la$-dependence observed
in Ref.\cite{Bordag2010} and in section 4.
  Note also that  detailed
 analysis and comparison for a number of  other papers are given
in Refs. \cite{BS2001}-\cite{BS2002a}.

 We have investigated the  dependence of the PT type on the value of coupling
 in the perturbation theory based on the  effective expansion parameter $N^{-1/3}$
 near the PT temperature. We found by calculating the contribution of the bubble
 chain graphs $D_{\phi}$ \Ref{Dphi} that
  this  change   is possible and this is the
  consequence of the
   sunset diagram contribution  near the $T_c$. It includes the $\la$ factor
  with the sigh which acts to influence the PT type.  Hence, we have to conclude that
  the problem on the PT type  must be considered with accounting for this value.
  This point can be investigated with the MC methods  on a lattice as in \cite{Bordag2010} for O(1) model.
   We left this problem for the future.

\end{document}